\documentclass[fleqn,10pt]{wlscirep}

\usepackage{color}

\definecolor{reddish}{HTML}{FBB4AE}
\definecolor{blueish}{HTML}{B3CDE3}

\usepackage{xcolor}
\usepackage[normalem]{ulem} 

\newcommand\red{\bgroup\markoverwith{\textcolor{reddish}{\rule[-.5ex]{2pt}{2.5ex}}}\ULon}
\newcommand\blue{\bgroup\markoverwith{\textcolor{blueish}{\rule[-.5ex]{2pt}{2.5ex}}}\ULon}
\newcommand\yellow{\bgroup\markoverwith{\textcolor{yellow}{\rule[-.5ex]{2pt}{2.5ex}}}\ULon}

\title{Spatio-temporal variations in the urban rhythm: the travelling waves of crime}

\author[1,*]{Marcos Oliveira}
\author[2]{Eraldo Ribeiro}
\author[3]{Carmelo Bastos-Filho}
\author[4]{Ronaldo Menezes}
\affil[1]{GESIS -- Leibniz Institute for the Social Sciences, Cologne, Germany}
\affil[2]{Computer Science, Florida Institute of Technology, Melbourne, USA}
\affil[3]{Escola Polit\'{e}cnica de Pernambuco, Universidade de Pernambuco, Recife, Brazil}
\affil[4]{Department of Computer Science, University of Exeter, UK}

\affil[*]{moliveira@biocomplexlab.org}

\begin{abstract}
In the last decades, the notion that cities are in a state of equilibrium with a centralised organisation has given place to the viewpoint of cities in disequilibrium and organised from bottom to up. In this perspective, cities are evolving systems that exhibit emergent phenomena built from local decisions. While urban evolution promotes the emergence of positive social phenomena such as the formation of innovation hubs and the increase in cultural diversity, it also yields negative phenomena such as increases in criminal activity. Yet, we are still far from understanding the driving mechanisms of these phenomena. In particular, approaches to analyse urban phenomena are limited in scope by neglecting both temporal non-stationarity and spatial heterogeneity. In the case of criminal activity, we know for more than one century that crime peaks during specific times of the year, but the literature still fails to characterise the mobility of crime. Here we develop an approach to describe the spatial, temporal, and periodic variations in urban quantities. With crime data from 12 cities, we characterise how the periodicity of crime varies spatially across the city over time. We confirm one-year criminal cycles and show that this periodicity occurs unevenly across the city. These `waves of crime' keep travelling across the city: while cities have a stable number of regions with a circannual period, the regions exhibit non-stationary series. Our findings support the concept of cities in a constant change, influencing urban phenomena---in agreement with the notion of cities not in equilibrium.

\end{abstract}
\begin{document}

\flushbottom
\maketitle
\thispagestyle{empty}


\section*{Introduction}
Cities evolve and undergo constant re-organisation as their population grow\cite{Batty1995,Batty2008}. 
This evolving process makes cities resilient and adaptive but also poses a challenge to analyse urban phenomena\cite{Southworth1993,Porta2014,Venerandi2017}. In recent years, evidence of nonlinear growth in urban indicators (e.g.,~wages, serious crime) with population size has motivated researchers to see cities as {\em complex systems}\cite{Batty2013,Bettencourt2010a,Gomez-Lievano2012,Bettencourt2007}.  
In this viewpoint, population increase combined with a high degree of social interaction in urban centres drive the emergence of assets such as innovation and wealth which emerge alongside drawbacks such as crime\cite{Bettencourt2007,Bettencourt2013,Pan2013,Gomez-Lievano2016}. 
Even though this perspective recognises cities continuously changing over time, we lack adequate approaches to study cities and better understand urban phenomena. 
In the case of crime, our knowledge of its mobility remains deficient\cite{DOrsogna2015}.
Crime's negative effects cannot be understated; not only does crime affect the lives of people but also the economy of cities. In the U.K. alone, criminal activity inflicts economic losses of some \textsterling124 billion yearly, a figure representing 7.7\% of that country's GDP\cite{peaceuk}.  Such impactful social and economic losses make crime reduction a key goal of societies worldwide\cite{Kates2003}. 

Successful policy-making relies on the understanding of the intricate dynamics of  cities. In general, researchers consider urban phenomena emergence as a result of population growth and the interaction among inhabitants\cite{Batty2013,Bettencourt2007,Bettencourt2010a,Gomez-Lievano2012,Bettencourt2013,Pan2013,Gomez-Lievano2016}. 
The characterisation of cities as complex systems finds quantitative support on the occurrence of regularities shown by power-law relationships associating indicators of urban growth. Many urban indicators have been shown to scale with population size $N$ according to the 
power law
${\smash Y \propto N^\beta}$, 
where the exponent $\beta$ relates to the indicator of interest\cite{Bettencourt2007,Gomez-Lievano2016}. For instance, indicators of infrastructure development such as length of electrical cables and roads display sub-linear scaling (i.e., ${\smash \beta < 1}$), whereas socio-economic indicators such as wages and number of patents display super-linear scaling (i.e., ${\smash \beta > 1}$). From this standpoint, researchers have also unveiled scaling regularities in criminal activity\cite{Bettencourt2007,Bettencourt2010,Gomez-Lievano2012,Alves2013,Hanley2016,Oliveira2017}. 
The increase of serious crime with population has been shown to be super-linear with ${\smash \beta \approx 1.16}$ in the U.S.\cite{Bettencourt2007}, and evidence has  revealed super-linearity in different types of crime and countries\cite{Gomez-Lievano2012,Alves2013,Hanley2016}.
Though many socio-economic factors influence crime\cite{Gordon2010}, the existence of scaling indicates a general mechanism underlying urban growth and suggests the presence of general regularities in cities, regardless of cities' particularities\cite{Bettencourt2013hyp}. 

In addition to displaying scaling regularity, urban social interactions are known to exhibit temporal regularity or {\em rhythms} of activity\cite{Batty2010,Morales2017}. These rhythms appear to result from the interplay between human circadian and the mobility patterns of city inhabitants\cite{Lenormand2015}. The daily human flux across the city defines areas of characteristic temporal rhythm\cite{Yuan2012}. While these  rhythms can be seen as {\em signatures} that characterise city regions, they also vary over time\cite{Lu2016}.  These variations can result from changes in human dynamics that occur at local and global levels of the city (e.g., an influx of new residents, closing establishments, new subway stations). 
 
Fluctuations in human dynamics can also affect temporal regularities of criminal activity, leading to periodic variations in crime rates such as annually or weekly.  In the case of annual seasonality, early studies linked crime to weather variations and aggression: heat stresses people, making aggressive people more likely to offend\cite{quetelet1842treatise,Cheatwood2009}. This weather-crime link has been since replaced by an understanding of relevant indirect effects\cite{Cheatwood2009}. For instance, weather affects social interaction which in turn affects crime. From this perspective, many researchers explain crime periodicity as the result of periodic changes in three requirements for crime: {\em offender}, {\em target}, and {\em opportunity}\cite{Cohen1979,Cheatwood2009}. For example, these three components may converge due to a seasonal increase of targets (e.g., more people outdoors during summer), thus creating crime seasonality. The same theoretic framework also explains other  crime periodicities according to the hour of the day\cite{Tranter1985}, the day of the week\cite{Anderson1984}, and the existence of holidays\cite{Lester1979}. 
The literature, however,  fails to account the location (i.e.,~spatial heterogeneity) and the continuous changes in cities (i.e.,~non-stationarity).

To model the temporal regularity of crime, most approaches in the literature use time-series analysis and its various tools such as  spectral analysis\cite{McPheters1974,Biermann2009,Breetzke2016,Venturini2016,Cohn2017}, spatial correlation\cite{Felice2014}, regression analysis\cite{Harries1983,Bollen1983,Warren1983,Cohn1993,LANDAU1993,Maes1994,VANKOPPEN1999,COHN2000,Cohn2003,Yan2004,Ceccato2005,Cusimano2010,McDowall2012,Carbone-Lopez2013,Tompson2013,Tompson2015,Santos2015,Linning2016,Linning2017}, cross correlation\cite{Toole2011}, and spatial point pattern tests\cite{Andresen2013,Linning2015a,Andresen2015}. These approaches assume a  temporal regularity of crime activity limited within fixed regional localities.  In these works, crime regularities have been shown to exist in city-level and local-level\cite{Sorg2011,Andresen2013,Andresen2015,Pereira2016,DeMelo2017,Chohlas-Wood2015,Felson2015,Breetzke2012,Cohen2003,Linning2015a} time series under the common assumption that crime's temporal regularities are stationary (i.e., the covariance of the time series remains constant over time).  
This stationary assumption implies that the urban dynamics in all regions across the city remains constant. 

The stationary assumption of crime periodicity\cite{McPheters1974,Biermann2009,Breetzke2016,Venturini2016,Cohn2017,Felice2014,Harries1983,Bollen1983,Warren1983,Cohn1993,LANDAU1993,Maes1994,VANKOPPEN1999,COHN2000,Cohn2003,Yan2004,Ceccato2005,Cusimano2010,McDowall2012,Carbone-Lopez2013,Tompson2013,Tompson2015,Santos2015,Linning2016,Linning2017,Toole2011,Sorg2011,Andresen2013,Andresen2015,Pereira2016,DeMelo2017,Chohlas-Wood2015,Felson2015,Breetzke2012,Cohen2003,Linning2015a} neglects that cities organise themselves continuously\cite{Batty1995,Batty2008}, preventing researchers to understand {\em crime mobility} in cities.  Indeed, even though evidence of temporal regularities in crime traces back to the nineteenth century\cite{quetelet1842treatise}, we still do not know how crime moves within a city. 
In our work, we advance the understanding of crime mobility by uncovering the characteristics of crime dynamics in cities. 
Here we describe temporal regularities in crime and the way they vary over time and across the city. 
We are interested in the general characteristics occurring regardless of the particularities of each city such as cultural or socio-economic aspects.
To investigate such urban phenomenon, however, we need to approach the city by considering both non-stationarity and spatial heterogeneity. 

In this paper, we relax the stationary assumption of temporal regularity and develop an approach to describe spatio-temporal regularities in criminal activities. 
Our main goal is to describe how the rhythms of crime are distributed across the city and how they vary over time. In our study, we are interested in the periodic cycles of crime (i.e., temporal regularities), instead of non-periodic changes in crime\cite{Oliveira2017}. For this, we analysed geolocated crime data from $12$ cities in the frequency domain. To account for spatial heterogeneity, we divide each city $c$ into non-overlapping geographic regions from which we build a region-specific time series $Y^c_i$ using data of each region $i$. We analyse the rhythmic variations in each $Y^c_i$ using wavelet analysis which enables us to decompose the time series into its various representative periodicities (i.e., multi-scale analysis) and analyse how these periodicities vary over time (i.e., non-stationarity). 
This non-stationary approach allows us to characterise the dynamics of crime in cities at both city and local levels. 
In the {\em city-level} time series, our analysis shows the existence of stationary circannual rhythms in most of the cities. 
In the \textit{local-level} time series, however, we show that this one-year period occurs unevenly across the city and that these criminal waves move across the city. 
Our results show that while cities exhibit a similar number of regions with annual periodicity throughout the time series, this set of regions vary over time due to the city-wide non-stationarity. In addition to circannual cycles, we also characterise other local-level periodicities which together form a remarkable signature of urban criminal rhythms. 
Our work recognises cities as dynamic and constantly organising processes, an approach that can lead to better understand them, and promote adequate policies to improve them.

\begin{table}[b!]
    \centering
    \begin{tabular}{p{3cm} r c  | p{2.8cm} r c  } 
    \toprule
    City & Population & Period & City & Population & Period\\ 
    \midrule
    Atlanta/GA      & $447,841$    & 2009--2015 & Portland/OR      & $609,456$  & 2004--2014 \\
    Chicago/IL      & $2,695,598$  & 2001--2015 & Raleigh/NC       & $431,746 $ & 2005--2015 \\
    Hartford/CT     & $125,017 $   & 2005--2015 & San Francisco/CA & $837,442$  & 2003--2015 \\
    Kansas City/MO  & $467,007$    & 2009--2015 & Santa Monica/CA  & $92,472$   & 2006--2015 \\
    New York/NY     & $8,550,405 $ & 2006--2015 & Seattle/WA       & $652,405$  & 2008--2015 \\
    Philadelphia/PA & $1,567,442$  & 2006--2015 & St. Louis/MO     & $318,416$  & 2008--2015 \\
    \bottomrule
    \end{tabular}
  \caption{\label{table1_datasets}Official disaggregated data sets of offences from different locations in the U.S.}
\end{table}
 
\section*{Results}
We examined spatiotemporal variations in crime periodicity using official criminal records of thefts from $12$ cities from the United States, spanning from $7$ to $15$ years (Table \ref{table1_datasets}). We investigated the temporal regularities of crime both in the entire city (i.e., city-level aggregated time series) and within the city (i.e., time series from small spatial regions). For this, we preprocessed the data to remove long-term trend and to decrease  skewness (Methods), resulting in discrete sequences $Y = \left\{y\left(1\right), y\left(2\right), \hdots, y\left(N\right)\right\}$ with observations of uniform time step $\delta t$. In this work, we analyse weekly numbers of crime, hence $y(t)$ represents the processed number of thefts at the week $t$. For simplicity, we denote $Y^c$ as the city-level time series of city $c$, whereas $Y^c_i$ represents the time series of the region $i$ in the city $c$. Throughout the paper, we adopt the term `wave of crime' to refer to any periodic fluctuation (e.g.,~annual, biannual) that is statistically significant in a time series of crime. Our goal is to describe the waves of crime as well as their changes over time (i.e., non-stationarity) and across the city (i.e., spatial heterogeneity).

\begin{figure}[b!]
\centering
\includegraphics[width=6in]{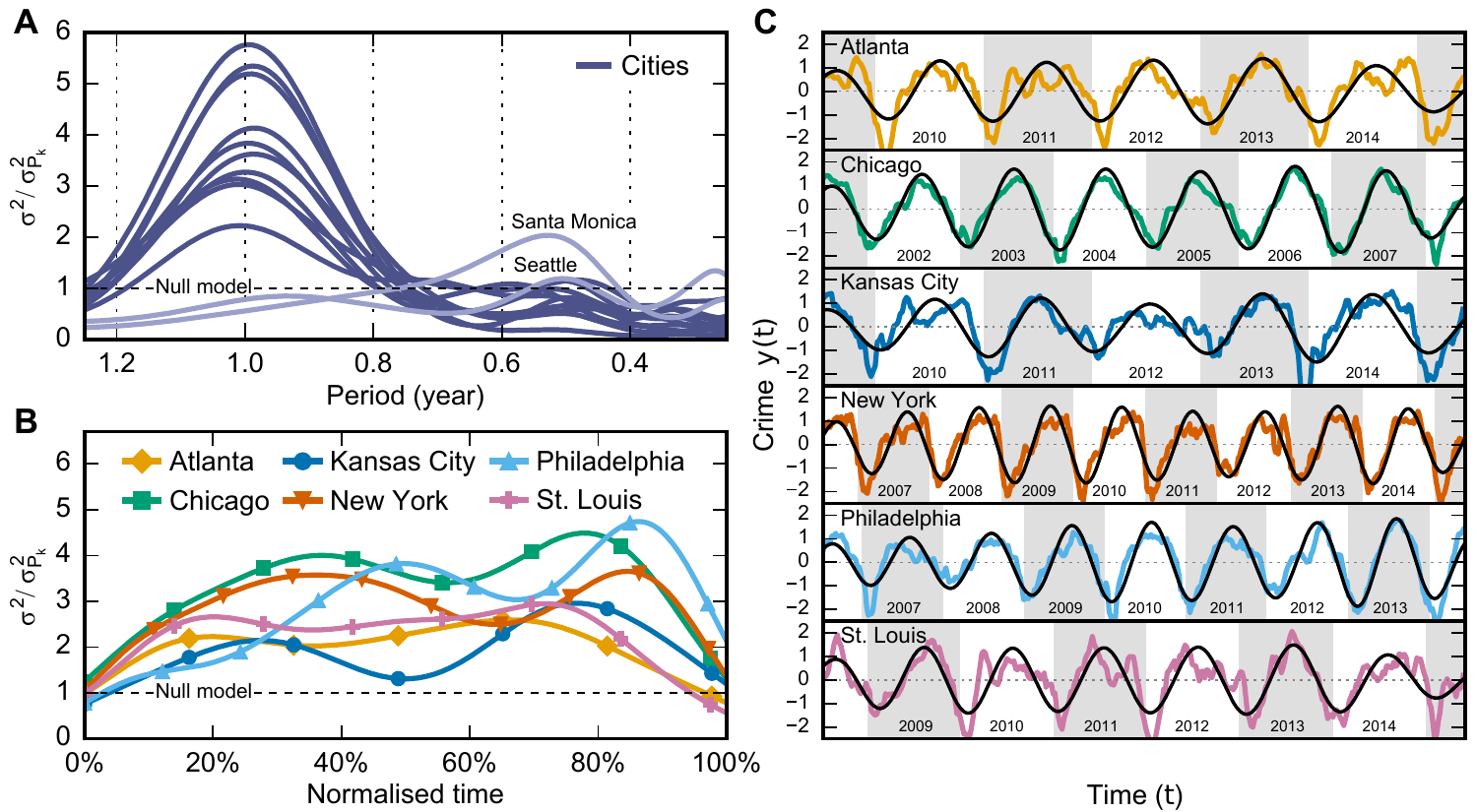}
\caption{Most of the analysed cities show a stationary circannual crime cycle. We estimated the criminal periodicity using the global wavelet spectrum (\textbf{A}) from the wavelet transform of the city-level time series (each solid curve), then compared against the null model $P_k$ generated from autocorrelated random noise (dashed line); only Santa Monica and Seattle failed to have the significant 1-year period (lighter colour). With the scale-averaged wavelet power (shown in \textbf{B} for selected cities), we found that these cycles are statistically stationary throughout the time series, except for Portland. Normalised time refers to the proportional amount of time to the whole time series. The temporal regularity can be seen in  the reconstructed waves of crime (\textbf{C}) using only the circannual band (black curves). }
\label{fig:fig1}
\end{figure}

\subsection*{Temporal regularities in cities}
We first characterised the temporal regularities (i.e., periodicity) of crime in the whole city in the frequency domain.  For this, we used the wavelet transform of the time series $Y^c$, defined as 
\begin{equation}
	{{\rm W}_{Y^c}(s,n)=\sqrt{\dfrac{\delta t}{s}}\sum_{t=1}^N y\left(t\right)\psi\left[\frac{\left(t-n\right)\delta t}{s}\right]},
\label{eq:discrete_wavelet}
\end{equation}
where $n \in \left\{1, 2, \ldots, N\right\}$ and $\psi$ is the wavelet function; we employed the Morlet wavelet (see Methods). The transform ${\rm W}_{Y^c}\left(s,n\right)$ gives us the contribution of a periodicity (scale) $s$ at different moments $n$ in the crime-activity time series. To identify cycles in the entire series, we calculated the average of ${\rm W}_{Y^c}\left(s,n\right)$ over $n$ (i.e., the global wavelet spectrum):
\begin{equation}
{\overline{{\rm W}}^2\left(s\right) = \frac{1}{N}\sum_{n=1}^{N}|{\rm W}_{Y^c}\left(s,n\right)|^2},
\label{eq:temporal_averaged_power}
\end{equation}
which gives us an unbiased estimate of the true power spectrum, providing an averaged picture of the periods in the time series\cite{Percival1995}. As shown in Fig. \ref{fig:fig1}A, the global spectrum enables us to test periodicities in crime against a null model (see Methods). With this approach, we confirmed previously documented evidence that crime exhibits seasonality\cite{BAUMER1996}. We found that most of the considered cities show a circannual wave of crime; only Santa Monica and Seattle failed to show annual periodicity, though both showed semestral periodicity.  

The global spectrum in equation (\ref{eq:temporal_averaged_power}) describes, however, only the expected periodic components in the time series and overlooks their possible non-stationarity. Yet, we wanted to evaluate the expected contribution of a periodic component at given time $n$. For this, we used the scale-averaged wavelet power, defined as 
\begin{equation}
{\overline{{\rm W}}^2_{j_1, j_2}\left(n\right) = \frac{\delta j \delta t}{C_\delta}\sum_{j=j_1}^{j_2}\frac{|{\rm W}_{Y^c}\left(s_j,n\right)|^2}{s_j}}, 
\label{eq:scale_averaged_power}
\end{equation}
which enables us to analyse the temporal evolution of a periodic signal in terms of a given band $b=\left(j_1, j_2\right)$. To examine the circannual stationarity, we evaluated the scale-averaged wavelet power with $j_1=0.8$ and $j_2=1.1$ (i.e., the circannual band) for each city and tested this periodicity against the null model. 
From the cities with the circannual periodicity, all of them displayed stationarity in the time series, except for Portland which displayed a weak non-stationarity; the circannual component is shown in Fig.~\ref{fig:fig1}B and Fig.~\ref{fig:fig1}C for some cities. 
In such city-level viewpoint, crime has a striking temporal regularity that may offer the impression of cities in equilibrium. We still neglect, however, spatial heterogeneity. With the continuous organisation process in cities, we expect local-level dynamics to change across the city, influencing urban phenomena such as crime, and thus leading to far-from-equilibrium cities\cite{Batty2012}.

\begin{figure}[b!]
\centering
\includegraphics[width=5.5in]{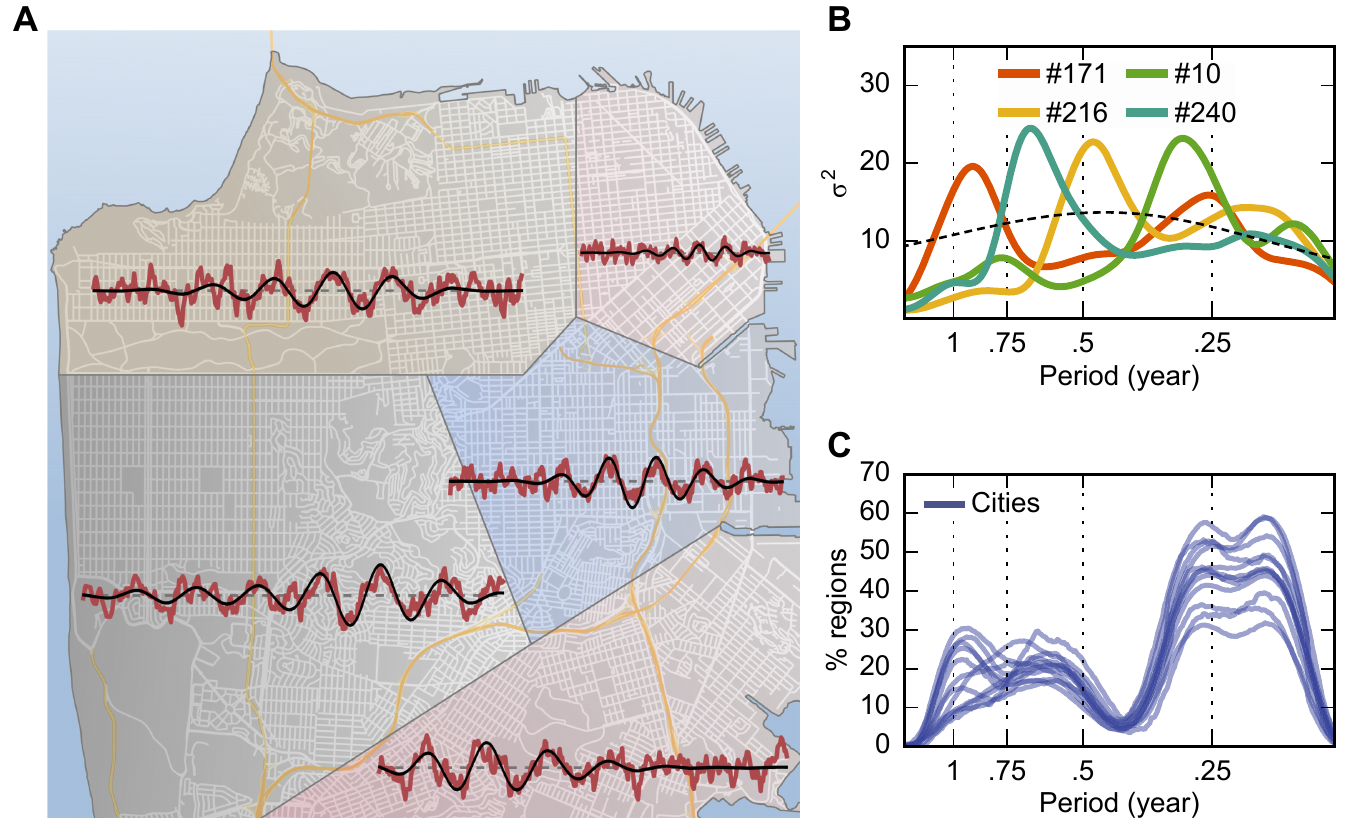}
\caption{Crime exhibits temporal regularities that are hidden in city-level aggregated data. To analyse the heterogeneity in cities, we divided each city into regions of same population size and created the time series $Y^c_i$ of each region. A pictorial example of San Francisco is shown in \textbf{A}. With the global wavelet spectrum of the wavelet transform of $Y^c_i$, we found that regions in the city might have other rhythms of crime, unseen in the aggregated data. For example, \textbf{B} depicts the global wavelet spectrum of selected regions (solid curves) in San Francisco and the null model (dashed curve). To describe these periods in the whole city, we developed the composed spectra $C_c(s)$ that gives us the proportion of regions that exhibit a period $s$. Each curve in \textbf{C} represents $C_c(s)$ of each considered city $c$ in our analysis, showing the signature of criminal periodicity in cities.}
\label{fig:fig2}
\end{figure}

\subsubsection*{Seasonality of crime in regions of the cities}
To verify whether the regularities found in $Y^c$ also occurs across the city, we now focus on the time series from smaller spatial aggregation units. We divided each city $c$ into regions of similar population size and built the time series $Y^c_i$ for each region $i$ (see Methods and an illustrative example in Fig. \ref{fig:fig2}A).
With this local-level viewpoint, we characterised the periods present in the regions of the city using the global wavelet spectrum from the wavelet transform of each $Y^c_i$. For instance, the spectrum of some regions in San Francisco can be seen in Fig.~\ref{fig:fig2}B, showing that regions have other crime rhythms that are hidden in city-level aggregated data. 
The global wavelet spectrum of the wavelet transform of $Y^c_i$ gives us a \textit{local-level} perspective of crime periodicity. 
We used this approach to characterise crime dynamics in cities while accounting for the heterogeneity in the dynamics across the regions of the city. 
For this, we developed the \emph{composed spectra}~$C_c\left(s\right)$ that describes a city $c$ in terms of the regions exhibiting time series with statistically significant global spectrum at each period~$s$. 
To build $C_c\left(s\right)$, we counted the number of regions $N_c\left(s\right)$ with significant period $s$, then divided $N_c\left(s\right)$ by the total number of regions in the city. 
The composed spectra $C_c\left(s\right)$ can be seen as the spectrum of the city $c$ which emerges from the bottom to up (i.e., from regions to the city), giving us a holistic view of the city.   

For each city $c$, we calculated $C_c(s)$ and found a signature of criminal periodicity (see Fig.~\ref{fig:fig2}C) in cities. Despite cities solely having a 1-year period in city-level analyses, regions in the city showed distinct periods that yield a remarkable signature of periods in the city as a whole. The city-level data hides the local crime dynamics that are present  in fine-grained levels of analysis. Though circannual cycles in city-level data, our bottom-up perspective depicts cities with a wide spectrum of dynamics that are embedded at the local level. From this signature of crime, we found that a 3-month period was the most prominent criminal cycle among the regions for all cities. 
The composed spectra are, however, an averaged picture of the city; changes in the rhythms may occur over time, and they must be taken into account.

\subsection*{Travelling waves of crime}
\begin{figure}[h!]
\centering
\includegraphics[width=5.8in]{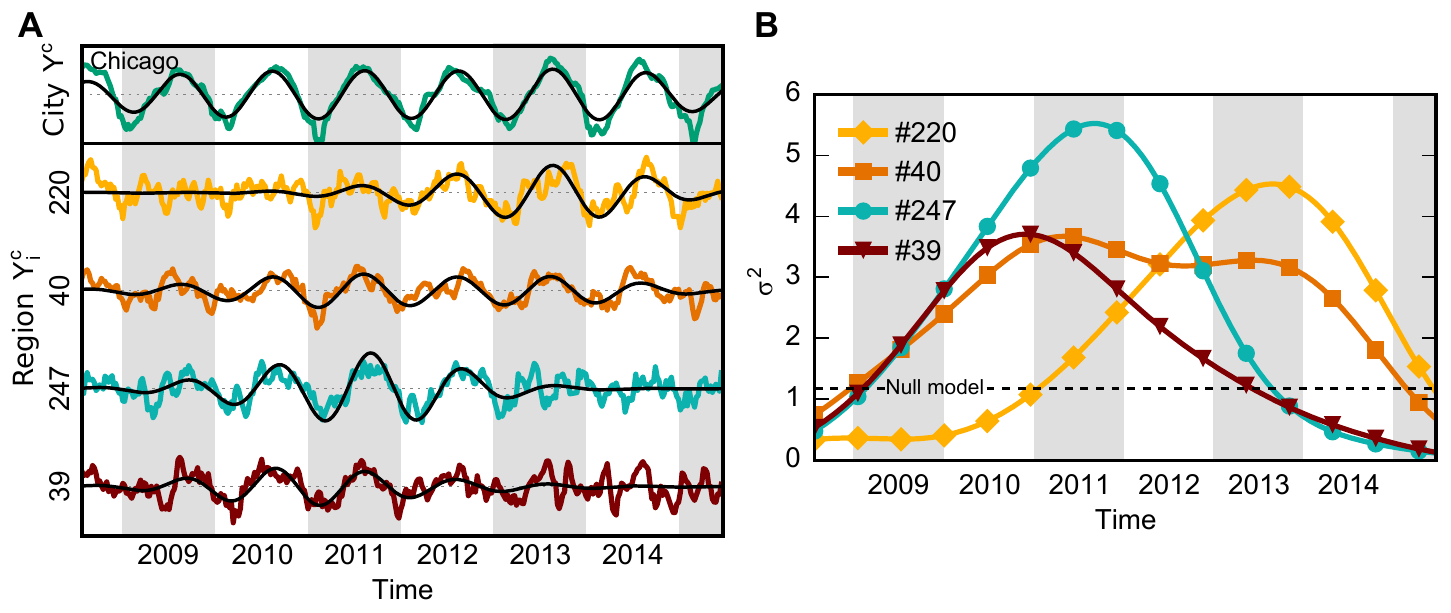}
\caption{The circannual waves of crime are non-stationary at local level. In the plots, four selected regions of Chicago, IL. Though the circannual period is stationary at city-level, (\textbf{A})  the periodicity changes throughout the time series. To track the non-stationarity in the regions, we used (\textbf{B}) the scale-averaged wavelet power of the wavelet transform of $Y^c_i$. }
\label{fig:fig3}
\end{figure}

To examine the stationarity in the regions of the city, we used the scale-averaged power $\smash{\overline{W}^2_{j_1, j_2}(n)}$ from the wavelet transform of each $Y^c_i$. In Fig.~\ref{fig:fig3}, we can see the non-stationarity of three regions in Chicago regarding the circannual component;~though region {\tt \#40} presents stationarity, region {\tt \#220} only starts to exhibit the circannual period in 2011, while region {\tt \#247} loses such periodicity in 2014. Despite the stationarity found in city-level time series, criminal waves at the local level change over time. Indeed, the apparent equilibrium seen in Fig.~\ref{fig:fig1} hides the constant changes taking place across the city. To examine these dynamics for the whole city, we analysed the number of regions that significantly show a given period at each time step. For this, we built the \textit{composed scale-averaged power} $C_c^b(t)$, defined as the number of regions that exhibit a statistically significant band $b=(j_1, j_2)$ at the time step~$t$ in the city $c$. The composed scale-averaged power enables us to analyse the whole city with respect to its dynamics on a given periodicity $b$. 

\begin{figure}[t!]
\centering
\includegraphics[width=5.7in]{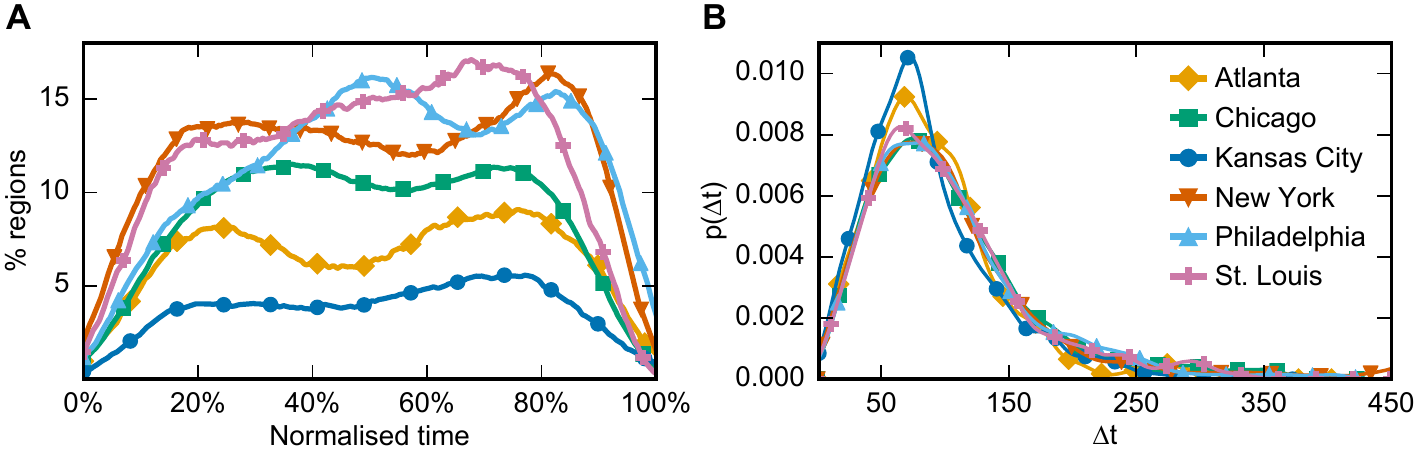}
\caption{The circannual waves of crime travel across the city. 
To examine the local-level waves of crime from a global perspective, we count the proportional number of regions that exhibit a statistically significant band $b$ at the time step $t$ in each city (i.e.,~the composed scale-averaged power ${\smash C_c^b(t)}$). In the case of the circannual band,  ${\smash C_c^b(t)}$ presents a typical value without much variability throughout the time series, suggesting a possible equilibrium at city level (shown in \textbf{A} for selected cities). However, the number of weeks $\Delta t$ that each region keeps this period continuously follows a probability distribution (\textbf{B}) that decays far before the complete criminal series, implying that these waves move across the city. Such continuous movement supports the view of cities in disequilibrium. In the plots, the legend in \textbf{B} also applies to the curves in \textbf{A}. Normalised time refers to the proportional amount of time to the whole time series. }
\label{fig:fig4}
\end{figure}

In this analysis, we focused on the 1-year rhythm of crime due to its extensive empirical evidence\cite{McPheters1974,Biermann2009,Breetzke2016,Venturini2016,Cohn2017,Felice2014,Harries1983,Bollen1983,Warren1983,Cohn1993,LANDAU1993,Maes1994,VANKOPPEN1999,COHN2000,Cohn2003,Yan2004,Ceccato2005,Cusimano2010,McDowall2012,Carbone-Lopez2013,Tompson2013,Tompson2015,Santos2015,Linning2016,Linning2017,Toole2011,Linning2015a,Sorg2011,Andresen2013,Andresen2015,Pereira2016,DeMelo2017,Chohlas-Wood2015,Felson2015,Breetzke2012,Cohen2003}. 
For all cities, we calculated $C_c^b(t)$ throughout the time series with respect to the circannual band.  We found that $C_c^b(t)$ exhibits a typical value without much variability over time for each city $c$ (Fig.~\ref{fig:fig4}A).  This result implies that cities exhibit a similar number of regions with 1-year cycle over time. Such finding is intriguing given the non-stationarity that we detected in the criminal waves. Despite regions having non-stationary time series, the number of regions with the circannual wave remains fairly the same throughout the series---a result that could suggest cities in equilibrium. 
Still, the composed scale-averaged power $C_c^b(t)$ only accounts for the \textit{number} of regions and neglects the regions themselves.  
Though the cities exhibit a similar number of regions with a circannual periodicity, $C_c^b(t)$ does not say whether this occurs because of the \textit{same set} of regions. 

\subsubsection*{Circannual periodicity moving across the city}
Ultimately, we wanted to know if the waves of crime move across the city. 
To characterise such mobility of crime, we are interested on the random variable $\Delta t^c_b$, defined as the amount of time that a $Y^c_i$ (i.e., a region in the city $c$) exhibits a significant periodicity with respect to the band $b$. Precisely, we counted the number of weeks that each region keeps the circannual band significant continuously.  In this analysis, we admit that regions may exhibit a criminal wave in distinct moments throughout the time series, so we measured the amount of time $Y^c_i$ presents significant periodicity independently.

For each city, we measured $\Delta t^c_b$ using the circannual band, and we found that their probability distributions decay much earlier than the total time of the criminal series (Fig.~\ref{fig:fig4}B). For each $\Delta t^c_b$, we selected the best model that described its distribution, and we found that the distribution of $\Delta t_b$ can be approximately described with a stretched exponential distribution (see Supplementary Material). This result indicates that, in general, the amount of time that a region presents a period is usually shorter than the whole time series. Not only do most of the regions exhibit non-stationarity, but also the criminal waves keep moving across the city.
Indeed, the distribution of $\Delta t^c_b$ coupled with the form of $C_c^b(t)$ implies waves of crime moving across the city. The presence of this temporal regularity in a region might indicate a place where crime occurs normally and a wave leaving a place suggests modifications in the dynamics of the area (e.g., closing establishments, deterioration of streets) disrupting the normalcy of crime. 

\section*{Discussion}
Cities are evolving systems that exhibit emergent phenomena built from local decisions, presenting messy but ordered patterns across different scales\cite{Batty2008}. 
Such evolving development makes cities to be in a continuous process of organisation. 
From this standpoint, we analysed crime---a severe threat to cities---and found agreement with the notion of cities not in equilibrium.
For this, we developed an approach to describe spatiotemporal variations in the rhythms of urban quantities using wavelet analysis.  
Our findings support the concept of cities in a constant change influencing urban phenomena. We confirmed the well-documented circannual rhythms of crime in cities, but we found that not only do these waves of crime occur unevenly at the local level, but they are also continuously travelling across the city. 

In our study, we were able to characterise general features in crime mobility. We analysed different cities and found remarkable regularities in crime dynamics, despite cultural and socio-economic differences between the cities. We described regularities in both city-level and local-level dynamics of crime, providing statistical characteristics for crime modelling. Though the proposal of a generative mechanism is beyond the scope of this paper, our work brings a new piece to the puzzle, alongside with other regularities in crime such as scaling\cite{Bettencourt2007} and  concentration\cite{Oliveira2017,PrietoCuriel2017}. 
Further investigations to understand the emergence of the temporal regularities might examine the specific cases of Santa Monica and Seattle (i.e., cities that fail to exhibit annual regularity) and also examine different spatial units of aggregation such as dynamic ones. 
One should not conclude that here we attempted to investigate what leads to crime (i.e., criminal aetiology), but rather we are interested in the statistical characteristics of crime when it takes place in cities rates. Though our study supports cities changing continuously, here we are not showing explicit changes in other urban factors besides crime rates. In fact, further efforts are needed to assess the relationship between aspects of the urban fabric (e.g., socio-economic factors\cite{Gordon2010}) and the travelling waves of crime; yet, such analyses must deal with the lack of fine-grained data. 

Our findings suggest that cities continuously change over time and, as such, policy-making needs evolving approaches and a constant assessment of the city. 
In this scenario, policy-makers need tools and up-to-date data to assess the changes happening in cities.
We believe that policy-makers may take advantage of our approach to track variations in the urban dynamics over time.
With the proper tools, we can learn more about cities and help to improve them.

\section*{Methods}
\subsection*{Preprocessing data}
In our analysis of criminal time series,  we first preprocessed the raw data to (1)~decrease skewness in the data, (2)~remove trends, and (3)~decrease intra-month variance. We then built the raw time series $r(t)$ using the number of offences that occurred in a given week within a spatial unit of aggregation. We used weekly numbers because evidence shows that offences happen more frequently during the weekends than during the weekdays\cite{Hipp2004}. 
Therefore, we created $r(t)$ using seven-days time windows, which means that each data point $r(t')$ is the number of occurrences in week $t'$. Because crime data is usually skewed (e.g., crime repeats, crime sprees\cite{Farrell2015}),
researchers often use a log transformation to decrease the skewness,  thus we transformed the data points as $x\left(t\right) = \log_{10} \left[r\left(t\right) + 1\right]$, which also gives us an intuition of percentage change in crime over time\cite{Hipp2004,Cohn2003,COHN2000}.
The time series of crime might exhibit a temporal trend that represents the long-term tendency of increase or decrease in the number of offences in a city over the years. In our analysis, however, we were interested on the periodicity around the trend and thus we removed the trends from the criminal series. For this, we first used the moving average of a series, defined as $\smash{{\rm M}^{n_1, n_2}\left[x\left(t\right)\right] = \frac{1}{n_2 - n_1}\sum_{n=n_1}^{n_2} x\left(t+n\right)}$,
to determine the long-term tendency in the series, using $n_1 = -26$ and $n_2=26$ (i.e., one year). Then, we removed the trend from the series as  $\smash{d\left(t\right) = x\left(t\right) - {\rm M}^{-26,26}\left[x\left(t\right)\right]}$, thus $d\left(t\right)$ consists of the detrended time series of crime in a city. 
Finally, we wanted to examine the criminal rhythms that are higher than one-month period; however, the high variance between weeks in each month might hide such tendencies in the series. Hence, we applied the moving-average filter with window size equals to $5$ to remove any intra-week dynamics, that is, $\smash{y\left(t\right) = {\rm M}^{0, 5}\left[d\left(t\right)\right]}$ (see also Supplementary Material).

\subsection*{Wavelet analysis}
We wanted to examine the temporal regularities in the criminal time series and investigate the temporal evolution of these regularities in the series. For this task, we used wavelet analysis to track the periodicity of a series over time. With this approach, we can evaluate non-stationarity in the series.  
Such analysis decomposes the time series using functions, called wavelets, that dilate (scale) to capture different frequencies and that translate (shift) in time to include changes \textit{with} time. Be $y\left(t\right)$ a time series, we can define the continuous wavelet transform of $y\left(t\right)$ with respect to the wavelet function $\psi$ as follows: 
\begin{equation}
{{\rm W}_y\left(s,\tau\right)=\frac{1}{\sqrt{s}}\int_{-\infty}^\infty x\left(t\right)\psi^{*}\left(\frac{t-\tau}{s}\right)\textnormal{d}t=\int_{-\infty}^\infty y\left(t\right)\psi^{*}_{s,\tau}(t)\textnormal{d}t},
\end{equation}
where ‘$*$’ denotes the complex conjugate. Because our data are discrete sequences, we used the definition described in Eq.~\ref{eq:discrete_wavelet} for our work\cite{Torrence1998}. The wavelet transform can be seen as the cross-correlation between the time series $x\left(t\right)$ and a set of functions $\psi^{*}_{s,\tau}\left(t\right)$, distributed over $t$, with different widths\cite{Cazelles2007}. For our analysis, we used the Morlet complex wavelet, defined as $\smash{\psi\left(t\right)=\pi^{-1/4}e^{i\omega_0t}e^{-t^2/2}}$, which yields a good balance between time and frequency localisation, and that has been used in time-series analysis\cite{Grenfell2001,Cazelles2008}. To examine the temporal evolution of the series, we used the \emph{local wavelet spectrum} as a tool to evaluate the periodicity in crime. The local wavelet spectrum is defined as 
$|{\rm W}(s, n)|^2$ and we can average it across both time and period (scale). Averaging the local spectrum across time, as defined in Eq.~\ref{eq:temporal_averaged_power}, yields the  \emph{global wavelet spectrum}. Averaging it across scale yields the \emph{scale-averaged wavelet power}, as described in Eq.~\ref{eq:scale_averaged_power}, and we analyse the temporal evolution of a periodic signal. In Eq.~\ref{eq:scale_averaged_power}, $C_\delta$ is the reconstruction factor defined for each wavelet function calculated by reconstructing a delta function $\delta$ from its wavelet transform; since we use the Morlet wavelet with $\omega_0=6$, the reconstruction factor $C_\delta=0.776$\cite{Torrence1998}. Here, we followed Torrence and Compo and used $s_j=s_02^{j\delta j}$ for $j=\{0, 1, \ldots, J\}$, and $J=\delta j^{-1} \log_2 \left(N\delta j/s_0\right)$, where $s_0$ is the smallest resolvable scale\cite{Torrence1998}. Note that since we analyse finite data, we must mind the borders because we do not have data beyond the bounds $\left[0, N\right]$. Thus, $\smash{W_X\left(s,n\right)}$ becomes unreliable as $n$ reaches the bounds---the cone of influence\cite{Torrence1998}. We treated the borders using the $e$-folding time of $\psi\left(\cdot\right)$ at each scale $s$, which for the Morlet wavelet is $\sqrt{2}s$. In practice, we padded the time series with zeroes and calculated the cone of influence.

The wavelet power spectrum gives us a measure of local variance. Its statistical significance was also of our interest. We used the method developed by Torrence and Compo\cite{Torrence1998}, which tests the wavelet power against a null model that generates a background power spectrum $P_k$. The test is given by: 
\begin{align}
	{D\left(\frac{|\overline{W}_X\left(s, n\right)|^2}
    {\sigma_X^2} < p\right) = \frac{1}{2}P_k\chi^2_\nu,}
	\label{TorrenceCompoTestPowerSpectrum}
\end{align}
where $\nu=2$ for complex wavelets (our case) and $\nu=1$ for real-valued wavelets\cite{Torrence1998}.

\subsection*{Creating spatial units}
We examined the time series of crime occurring in small spatial units across a city. For this, we constructed these units using a method developed by Oliveira et al.\cite{Oliveira2017} which splits a city into regions of similar resident population. To choose the number of regions for each city, first we investigated the relationship between the crime rate of the regions and the number of regions to analyse. We counted the number of regions $R_c^\varphi\left(r\right)$ where crime rate is higher than $\varphi$ when the city $c$ is composed of $r$ regions with same population size. In all cities, using $\varphi = 1.0$, we found that $R_c^\varphi\left(r\right)$ increases with $r$ until it reaches a maximum value $R_c^\varphi\left(r^u_c\right)$ of regions with crime rate higher than $1$ crime per week. This result was expected because crime is unevenly distributed\cite{Oliveira2017}. As we divide a city into an increasing number of regions, the spatial units have smaller area which implies lower probability that offences occur in the same unit. Nevertheless, as we wanted to assess the waves of crime across the city, our analysis needed a number of spatial units that cover the city and that have enough data points. The requirement for sufficient amount of data in each region is controlled by the threshold $\varphi$: high values lead to take into account regions where few offences were recorded. The city-coverage requirement relates to the number of regions included in the analysis. Though more regions imply smaller regions, more regions allow to track criminal waves across different places. In our analysis of small spatial units, we define the average of $1$ crime per week as the minimum amount of crime to analyse a place, splitting each city $c$ into $r^u_c$ regions while setting $\varphi=1.0$ (see Supplementary Material).

\subsection*{Data Availability}
All crime data are official open data sets that are available as described in the Supplementary Material file.

\bibliography{sample}


\section*{Acknowledgements}

Marcos Oliveira would like to thank the Science Without Borders program (CAPES, Brazil) for financial support under grant 1032/13-5. This work was partially supported by the Army Research Office under grant W911NF-17-1-0127-P00001. We are grateful for Christopher Torrence and Gilbert Compo for making their wavelet code available. We also thank Bernard Cazelles and Ottar Bj{\o}rnstand for making available their data sets which allowed us to validate some of our implementations. 

\section*{Author contributions statement}

M.O. designed the study and conducted the experiments. M.O., E.R., and R.M. analysed the results. M.O. and E.R. wrote the manuscript. All authors reviewed the manuscript.

\section*{Competing  interests}



The authors declare no competing interests as defined by Nature Research, or other interests that might be perceived to influence the results and/or discussion reported in this paper.

\end{document}